\documentclass[twocolumn,amsmath,amssymb,showpacs,superscriptaddress]{revtex4}
\usepackage[dvipdfmx]{graphicx}
\usepackage{graphicx}
\usepackage{dcolumn}
\usepackage{bm}
\usepackage{amsmath}
\usepackage{amsfonts}
\usepackage{amssymb}
\usepackage{color}
\makeatletter
\pacs{45.70.Qj, 47.11.Mn, 64.70.F-, 83.10.Mj}

\begin{document}
\title{Simulation of cohesive fine powders under a plane shear}
\author{Satoshi Takada}
\affiliation{Yukawa Institute for Theoretical Physics, Kyoto University, Kitashirakawa Oiwakecho, Sakyo-ku, Kyoto 606-8502, Japan}
\author{Kuniyasu Saitoh}
\affiliation{Faculty of Engineering Technology, MESA+, University of Twente, 7500 AE Enschede, The Netherlands}
\author{Hisao Hayakawa}
\affiliation{Yukawa Institute for Theoretical Physics, Kyoto University, Kitashirakawa Oiwakecho, Sakyo-ku, Kyoto 606-8502, Japan}

%%%%%%%%%%%%%%%%%%%%%%%%%%%%%%
\begin{abstract}
Three dimensional molecular dynamics simulations of cohesive dissipative powders under a plane shear are performed. 
We find the various phases depending on the dimensionless shear rate and the dissipation rate as well as the density.
We also find that the shape of clusters depends on the initial condition of velocities of particles when the dissipation is large.
Our simple stochastic model reproduces the non-Gaussian velocity distribution function appearing in the coexistence phase of a gas and a plate.
\end{abstract}
\maketitle

%%%%%%%%%%%%%%%%%%%%%%%%%%%%%%
\section{Introduction}
Fine powders, such as aerosols, volcanic ashes, flours, and toner particles are commonly observed in daily life.
The attractive interaction between fine powders plays major roles \cite{Krupp1967,Visser1989,Valverde2000, Tomas2001, Cansell2003,Tomas2004,Castellanos2005, Butt2005, Tykhoniuk2007,Tomas2007,Calvert2009,Israelachvili2011,Ommen2012},
while there are various studies discussing the effects of cohesive forces between macroscopic powders \cite{Derjaguin1956, Johnson1971, Derjaguin1975, Yen1992, Thornton1996, Hornbaker1997, Bocquet1998, Mikami1998, Nase2001, Israelachvili2011, Bartels2005,  Willett2000,Herminghaus2005, Mitarai2006, Royer2009, Ulrich2012,
Luding2003, Luding2005, Luding2008, Luding2014, Luding201407}.
For example, the Johnson-Kendall-Roberts (JKR) theory describes microscopic surface energy for the contact of cohesive grains \cite{Johnson1971, Israelachvili2011}.
The others study the attractive force caused by liquid bridge for wet granular particles \cite{Hornbaker1997, Bocquet1998, Willett2000, Nase2001, Herminghaus2005, Mitarai2006, Ulrich2012}.
It should be noted that the cohesive force cannot be ignored for small fine powders.
Indeed, the intermolecular attractive force always exists.
Moreover, the inelasticity plays important roles
when powders collide, because there are some excitations of internal vibrations, radiation of sounds, and deformations \cite{Gerl1999, Hayakawa2002, Awasthi2006, Brilliantov2007, Suri2008, Tanaka2012, Kuninaka2009, Kuninaka2012, Saitoh2010, Murakami2014}.
\\
\quad
Let us consider cohesive powders under a plane shear.
So far there exist many studies for one or two effects of the shear, an attractive force, and an inelastic collision \cite{Saitoh2007, Yasuoka1998, Lasinski2004, Conway2004, Goldhirsch1993_1, Goldhirsch1993_2,McNamara1993, McNamara1996, Brilliantov2004,Hansen1969, Nicolas1979, Adachi1988, Lotfi1992, Johnson1993, Kolafa1993, Heist1994, Laaksonen1994, Alam2013, Shukla2013}, 
but we only know one example for the study of the jamming transition to include all three effects \cite{Gu2014}.
On the other hand, when the Lennard-Jones (LJ) molecules are quenched below the coexistence curve of gas-liquid phases \cite{Hansen1969, Nicolas1979, Adachi1988, Lotfi1992, Johnson1993, Kolafa1993}, a phase ordering process proceeds after the nucleation takes place \cite{Heist1994, Laaksonen1994,Yasuoka1998}.
It is well known that clusters always appear in freely cooling processes of granular gases \cite{Goldhirsch1993_1, Goldhirsch1993_2,McNamara1996}.
Such clustering processes may be understood by a set of hydrodynamic equations of granular gases \cite{McNamara1993, Brilliantov2004}.
When we apply a shear to the granular gas, 
there exist various types of clusters such as 2D plug, 2D wave, or 3D wave for three dimensional systems \cite{Lasinski2004, Conway2004,Saitoh2007, Alam2013, Shukla2013}.
\\
\quad
In this paper, 
we try to characterize nonequilibrium pattern formation of cohesive fine powders 
under the plane shear by the three dimensional molecular dynamics (MD) simulations 
of the dissipative LJ molecules under the Lees-Edwards boundary condition \cite{Lees1972}.
In our previous paper \cite{Takada2013_2}, we have mainly focused on the effect of dissipation on the pattern formation in Sllod dynamics \cite{Evans1984, Evans2008}.
In this study, we systematically study it by scanning a large area of parameters space to draw the phase diagrams with respect to the density, the dimensionless shear rate, and the dissipation rate without the influence of Sllod dynamics.
\\
\quad
The organization of this paper is as follows.
In the next section, we introduce our model and setup for this study.
Section \ref{sec:Results}, the main part of this paper, is devoted to exhibit the results of our simulation.
In Sec.~\ref{sec:Phase}, we show the phase diagrams for several densities, 
each of which has various distinct steady phases.
We find that the system has a quasi particle-hole symmetry.
We also find that the steady states depend on the initial condition of velocities of particles when the dissipation is large.
In Sec.~\ref{sec:VDF}, we analyze the velocity distribution function, and try to reproduce it by solving the Kramers equation with Coulombic friction under the shear.
In the last section, we discuss and summarize our results.
In Appendix \ref{sec:LEbc}, we study the pattern formation of dissipative LJ system 
under the physical boundary condition.
In Appendix \ref{sec:Coulombic}, we illustrate the existence of Coulombic friction 
near the interface of the plate-gases coexistence phase.
In Appendix \ref{sec:av_pypz}, we demonstrate that the viscous heating term near the interface is always positive.
In Appendix \ref{sec:perturbation}, we present a perturbative solution of the Kramers equation.
In Appendix \ref{sec:moment}, we show the detailed calculations for each moment.
In Appendix \ref{sec:rotation_f}, we show the detailed calculations of the velocity distribution function.
%%%%%%%%%%%%%%%%%%%%%%%%%%%%%%
%%%%%%%%%%%%%%%%%%%%%%%%%%%%%%
%%%%%%%%%%%%%%%%%%%%%%%%%%%%%%
\section{Molecular dynamics simulation}
In this section, we explain our model and setup of MD 
for cohesive fine powders under a plane shear.
We introduce our model of cohesive fine powders in Sec.~\ref{sec:model} and explain our numerical setup in Sec.~\ref{sec:setup}.
%%%%%%%%%%%%%%%%%%%%%%%%%%%%%%
\begin{figure}[htbp]
	\includegraphics[width=55mm]{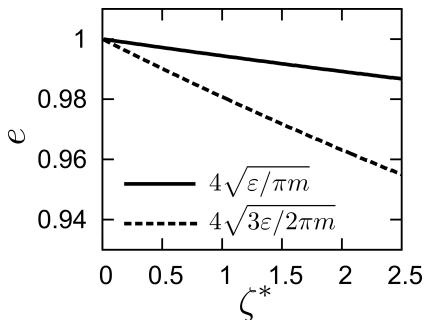}
	\caption{The relationship between the dimensionless dissipation rate $\zeta^*$ and the coefficient of restitution $e$ when the pre-collisional relative velocities (solid and dashed lines) are given by $4\sqrt{\varepsilon/\pi m}$ and $4\sqrt{3\varepsilon/2\pi m}$, respectively.}
	\label{fig:COR}
\end{figure}
%%%%%%%%%%%%%%%%%%%%%%%%%%%%%%
%%%%%%%%%%%%%%%%%%%%%%%%%%%%%%
\subsection{Model}\label{sec:model}
We assume that the interaction between two cohesive fine powders can be described by LJ potential, 
and an inelastic force caused by collisions with finite relative speeds.
The explicit expression of LJ potential is given by
\begin{equation}
U^{\rm LJ}(r_{ij})=
4\varepsilon\Theta(r_{\rm c}-r_{ij})\left[ \left( \frac{\sigma}{r_{ij}} \right)^{12} - \left(
\frac{\sigma}{r_{ij}} \right)^{6} \right]\label{eq:LJ}
\end{equation}
with a step function $\Theta(r)=1$ and $0$ for $r>0$ and $r\le0$, respectively,
where $\varepsilon$, $\sigma$, and $r_{ij}$ are the well depth, the diameter of the repulsive core,
and the distance between the particles $i$ and $j$, respectively.
Here, we have introduced the cutoff length $r_{\rm c}=3.0\sigma$ to save the computational cost,
i.e. $U^{\rm LJ}(r)=0$ for $r\ge r_{\rm c}$.
To model the inelastic interaction, we introduce a viscous force between colliding two particles as
\begin{equation}
\bm{F}^{\rm vis}(\bm{r}_{ij},\bm{v}_{ij})=-\zeta\Theta(\sigma-r_{ij})(\bm{v}_{ij} \cdot \hat{\bm{r}}_{ij})
\hat{\bm{r}}_{ij},\label{eq:dashpot}
\end{equation}
where $\zeta$, $\hat{\bm{r}}_{ij}\equiv \bm{r}_{ij}/r_{ij}$, and $\bm{v}_{ij}=\bm{v}_i-\bm{v}_j$ are the dissipation rate, 
a unit vector parallel to $\bm{r}_{ij}=\bm{r}_i-\bm{r}_j$, and the relative velocity between the particles, respectively.
Here, $\bm{r}_\alpha$ and $\bm{v}_\alpha$ ($\alpha=i, j$) are, respectively, the position and velocity of the particle.
It should be noted that the range of inelastic interaction is only limited within the distance $\sigma$.
From Eqs.~(\ref{eq:LJ}) and (\ref{eq:dashpot}), the force acting on the $i$-th particle is given by
\begin{equation}
\bm{F}_i=-\sum_{j\neq i}\bm\nabla_i U^{\rm LJ}(r_{ij})+\sum_{j\neq i}\bm{F}^{\rm vis}(\bm{r}_{ij},\bm{v}_{ij}).
\end{equation}
\quad
Our LJ model has an advantage to know the detailed properties in equilibrium \cite{Hansen1969, Nicolas1979, Adachi1988, Lotfi1992, Johnson1993, Kolafa1993}.
The normal restitution coefficient $e$, defined as a ratio of post-collisional speed to pre-collisional speed,
depends on both the dissipation rate $\zeta$ and incident speed.
For instance, the particles are nearly elastic, i.e. the restitution coefficient, $e=0.994$ for the case of $\zeta=\sqrt{\varepsilon/m\sigma^2}$ 
and the incident speed $\sqrt{\varepsilon/m}$,
where $m$ is the mass of each colliding particle.
Figure \ref{fig:COR} plots the restitution coefficient against the dimensionless dissipation rate $\zeta^\ast=\zeta\sqrt{m\sigma^2/\varepsilon}$, 
where the incident speeds are given by $4\sqrt{\varepsilon/\pi m}$ and $4\sqrt{3\varepsilon/2\pi m}$, respectively.
We restrict the dissipation rate to small values in the range $0<\zeta^\ast\le3.2$.
Note that small and not too large inelasticity is necessary to reproduce a steady coexistence phase  between a dense and a dilute region, 
which will be analyzed in details in this paper.
Indeed, the system cannot reach a steady state without inelasticity, while all particles are absorbed in a big cluster when inelasticity is large.
In this paper, we use three dimensionless parameters to characterize a system:
the dimensionless density $n^*=n\sigma^3=N\sigma^3/L^3$, 
the shear rate $\dot\gamma^\ast=\dot\gamma\sqrt{m\sigma^2/\varepsilon}$, and 
the dissipation rate $\zeta^\ast=\zeta\sqrt{m\sigma^2/\varepsilon}$.
It should be noted that the well depth $\varepsilon$ is absorbed in the dimensionless shear rate and the dissipation rate.
Thus, we may regard the control of two independent parameters as the change of the well depth.
%%%%%%%%%%%%%%%%%%%%%%%%%%%%%%
%%%%%%%%%%%%%%%%%%%%%%%%%%%%%%
%%%%%%%%%%%%%%%%%%%%%%%%%%%%%%
\begin{figure}[htbp]
	\includegraphics[width=60mm]{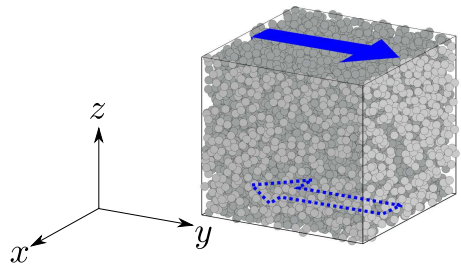}
	\caption{A snapshot of our simulation in a uniformly sheared state.
			We apply a plane shear in $xy$ plane, that is,
			we choose $y$-axis as the shear direction and $z$-axis as the velocity gradient direction.}
	\label{fig:setup}
\end{figure}
%%%%%%%%%%%%%%%%%%%%%%%%%%%%%%
\subsection{Setup}\label{sec:setup}
Figure \ref{fig:setup} is a snapshot of our MD for a uniformly sheared state,
where we randomly distribute $N=10^4$ particles in a cubic periodic box
and control the number density $n$ by adjusting the linear system size $L$.
At first, we equilibrate the system by performing the MD 
with the Weeks-Chandler-Andersen potential  \cite{Chandler1970, Weeks1971}
during a time interval $100\sqrt{m\sigma^2/\varepsilon}$.
We set the instance of the end of the initial equilibration process as the origin of the time for later discussion.
Then, we replace the interaction between the particles by the truncated LJ potential (\ref{eq:LJ}) 
with the dissipation force, Eq.~(\ref{eq:dashpot})
under the Lees-Edwards boundary condition.
As shown in Appendix \ref{sec:LEbc}, the results under the Lees-Edwards boundary condition are almost equivalent to 
those under the flat boundary.
The time evolution of position $\bm{r}_i=(x_i, y_i, z_i)$ is given by Newton's equation of motion
$md^2\bm{r}_i/dt^2=\bm{F}_i$.
%%%%%%%%%%%%%%%%%%%%%%%%%%%%%%
%%%%%%%%%%%%%%%%%%%%%%%%%%%%%%
%%%%%%%%%%%%%%%%%%%%%%%%%%%%%%
\section{Results}\label{sec:Results}
In this section, we present the results of our MD.
In Sec.~\ref{sec:Phase}, we draw phase diagrams of the spatial structures of cohesive fine powders.
In Sec.~\ref{sec:VDF}, we present the results of velocity distribution functions 
and reproduce it by solving a phenomenological model.
\subsection{Phase diagram}\label{sec:Phase}
%%%%%%%%%%%%%%%%%%%%%%%%%%%%%%
%%%%%%%%%%%%%%%%%%%%%%%%%%%%%%
\begin{table}[htbp]% add [H] placement to break table across pages
	\caption{The dimensionless parameters used in Fig.~\ref{fig:steady}.
	\label{tab:phase}}
	\begin{ruledtabular}
		\begin{tabular}{cccc}
			% Lines of table here ending with \\
			Phase & $n^\ast$ & $\dot\gamma^\ast$ & $\zeta^\ast$ \\ \hline
			(a) & $0.305$ & $10^{-1}$ & $10^{-2}$ \\
			(b) & $0.0904$ & $10^{-0.5}$ & $10^{0.5}$ \\
			(c) & $0.156$ & $10^{-0.5}$ & $10^{0}$ \\
			(d) & $0.305$ & $10^{-0.2}$ & $10^{0.2}$ \\
			(e) & $0.0904$ & $10^{-2}$ & $10^{-1}$ \\
			(f) & $0.156$ & $10^{-1}$ & $10^{-0.75}$ \\
			(g) & $0.305$ & $10^{-1}$ & $10^{-1}$ \\
			(h) & $0.723$ & $10^{-2}$ & $10^{-1}$ \\
			(i) & $0.723$ & $10^{-2}$ & $10^{-2}$ \\
		\end{tabular}
	\end{ruledtabular}
\end{table}
%%%%%%%%%%%%%%%%%%%%%%%%%%%%%%
%%%%%%%%%%%%%%%%%%%%%%%%%%%%%%
\begin{figure}[htbp]
	\includegraphics[width=84mm]{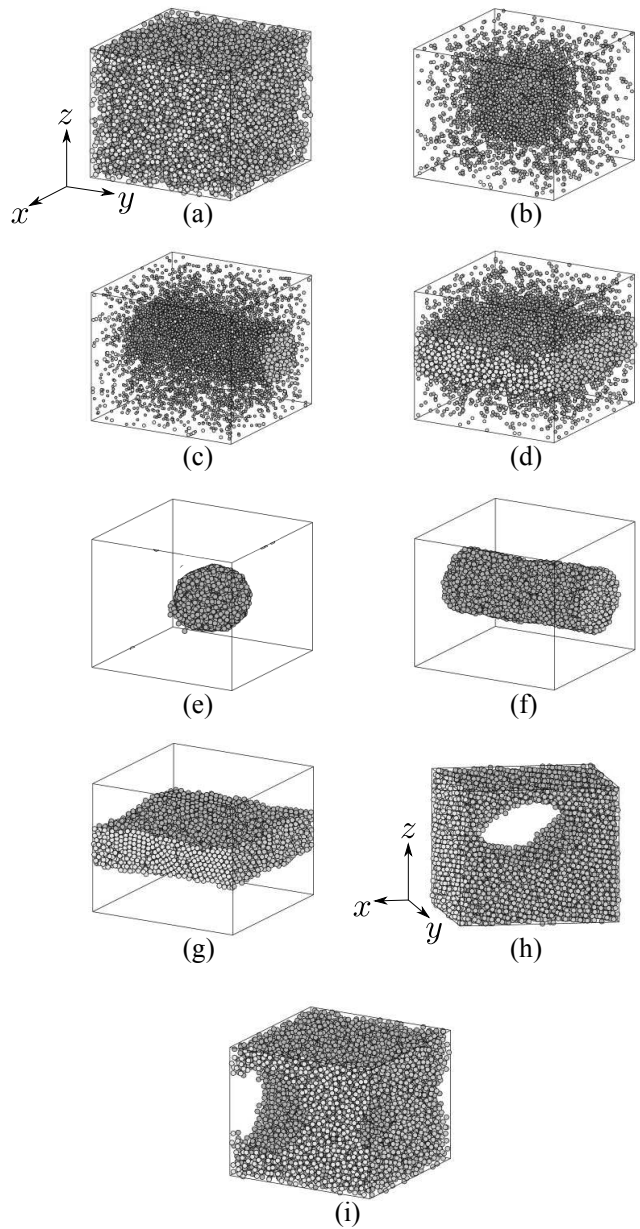}
	\caption{(Color online) Steady patterns made of the particles under the plane shear:
			(a) uniformly sheared phase,
			(b) coexistence of a spherical-droplet and gas,
			(c) coexistence of a dense-cylinder and gas,
			(d) coexistence of a dense-plate and gases,
			(e) an isolated spherical-droplet,
			(f) an isolated dense-cylinder,
			(g) an isolated dense-plate, 
			(h) an inverse cylinder, and
			(i) an inverse droplet, 
			where the corresponding dimensionless parameters $n^\ast$, $\dot\gamma^\ast$, 
			and $\zeta^\ast$ for (a)--(i) are listed in Table \ref{tab:phase}.
			We note that gas particles in (b), (c) and (d) 
			are drawn smaller than the real size for visibility.}
	\label{fig:steady}
\end{figure}
%%%%%%%%%%%%%%%%%%%%%%%%%%%%%%
%%%%%%%%%%%%%%%%%%%%%%%%%%%%%%
\begin{figure}[htbp]
	\includegraphics[width=85mm]{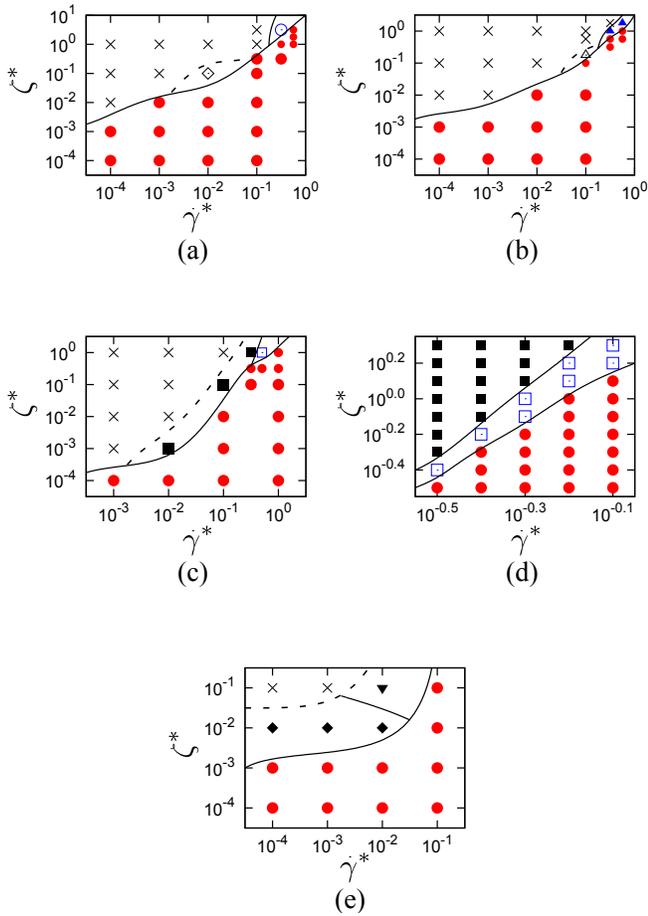}
	\caption{(Color online) Phase diagrams for various densities, where the dimensionless densities are given by
			(a) $n^*=0.0463$, (b) $0.156$, (c) $0.305$, 
			(d) $0.305$ for $10^{-0.5}\le \dot\gamma^* \le 10^{-0.1}$, and (e) $0.723$, respectively.
			The spatial patterns corresponding to Fig.~\ref{fig:steady}(a)--(i) are represented by
			red filled circles (Fig.~\ref{fig:steady}(a)), 
			blue open circles (Fig.~\ref{fig:steady}(b)), 
			blue filled upper triangles (Fig.~\ref{fig:steady}(c)), 
			blue open squares (Fig.~\ref{fig:steady}(d)), 
			black open diamond (Fig.~\ref{fig:steady}(e)), 
			black open upper triangles (Fig.~\ref{fig:steady}(f)),
			black filled squares (Fig.~\ref{fig:steady}(g)),
			black filled lower triangles (Fig.~\ref{fig:steady}(h)), and
			black filled triangles (Fig.~\ref{fig:steady}(i)), respectively.
			The steady states represented by the cross marks show various patterns depending on the initial velocities of particles.}
	\label{fig:phasediagram}
\end{figure}
%%%%%%%%%%%%%%%%%%%%%%%%%%%%%%
%%%%%%%%%%%%%%%%%%%%%%%%%%%%%%
\begin{figure}[htbp]
	\includegraphics[width=80mm]{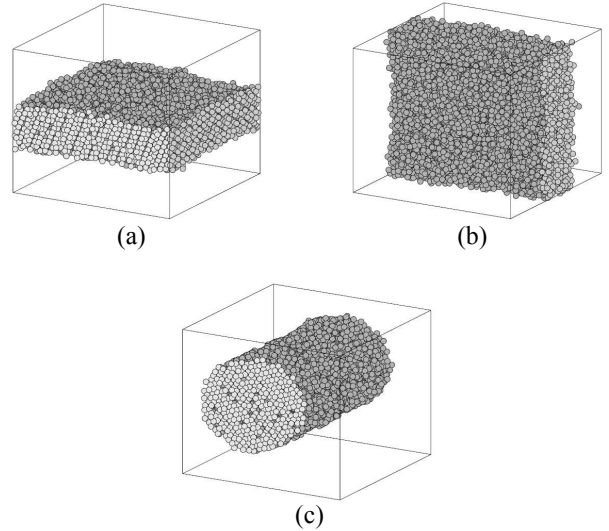}
	\caption{(Color online) Typical examples of initial configuration dependence when we start from the identical parameters ($n^*=0.305$, $\dot\gamma^*=10^{-3}$, $\zeta^*=10^{-2}$): 
	(a) a dense-plate cluster parallel to $xy$ plane, 
	(b) a dense-plate cluster parallel to $yz$ plane and 
	(c) a dense-cylinder cluster parallel to $x$-axis.}
	\label{fig:initial_configuration}
\end{figure}
%%%%%%%%%%%%%%%%%%%%%%%%%%%%%%
%%%%%%%%%%%%%%%%%%%%%%%%%%%%%%
\begin{figure}[htbp]
	\includegraphics[width=80mm]{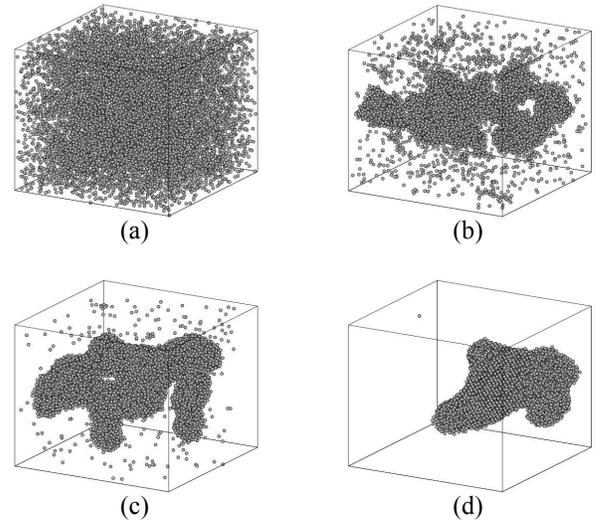}
	\caption{(Color online) Time evolution of configurations for $n^*=0.0904$, $\dot\gamma^*=10^{-1}$, $\zeta^*=10^{0.5}$.
(a) $t^*=0$, (b) $50$, (c) $100$, and (d) $550$.}
	\label{fig:dependence}
\end{figure}
%%%%%%%%%%%%%%%%%%%%%%%%%%%%%%
%%%%%%%%%%%%%%%%%%%%%%%%%%%%%%
%%%%%%%%%%%%%%%%%%%%%%%%%%%%%%
%%%%%%%%%%%%%%%%%%%%%%%%%%%%%%
%%%%%%%%%%%%%%%%%%%%%%%%%%%%%%
Figure \ref{fig:steady} displays typical patterns formed by the particles in their steady states,
which are characterized by the dimensionless parameters $n^*$, $\dot\gamma^\ast$, and $\zeta^\ast$ as listed in Table \ref{tab:phase}.
Figure \ref{fig:phasediagram} shows phase diagrams in the steady states for
(a) $n^\ast=0.0904$, (b) $0.156$, (c) $0.305$, and (d) $0.723$.
Three of these phases, Figs.~\ref{fig:steady}(a), (d) and (g), are similar to 
those observed in a quasi two-dimensional case with Sllod dynamics \cite{Takada2013}.
If the shear is dominant, the system remains in a \emph{uniformly sheared phase} (Fig.~\ref{fig:steady}(a)).
However, if the viscous heating by the shear is comparable with the energy dissipation, 
we find that a \emph{spherical-droplet}, a \emph{dense-cylinder}, and a \emph{dense-plate} coexist 
for extremely dilute ($n^\ast=0.0904$), dilute ($n^\ast=0.156$), and moderately dense ($n^\ast=0.305$) gases, respectively (Figs.~\ref{fig:steady}(b)--(d)).
These three coexistence phases are realized by the competition between the equilibrium phase transition and the dynamic instability
caused by inelastic collisions.
Furthermore, if the energy dissipation is dominant, there are no gas particles in steady states (Figs.~\ref{fig:steady}(e)--(g)).
For an extremely high density case ($n^\ast=0.723$), 
we observe an \emph{inverse-cylinder},
where the vacancy forms a ``hole" passing through the dense region 
along the $y$-axis (Fig.~\ref{fig:steady}(h)), 
and an \emph{inverse-droplet}, where the shape of the vacancy is spherical (Fig.~\ref{fig:steady}(i)).
In our simulation, the role of particles in a dilute system corresponds to that of vacancies in a dense system.
Thus, the system has a quasi particle-hole symmetry.
\\
\quad
Moreover, the shape of clusters depends on the initial condition of velocities of particles,
even though a set of parameters such as the density, the shear rate, the dissipation rate and the variance of the initial velocity distribution function are identical when the dissipation is strong.
We observe 
a dense-plate parallel to $xy$ plane (Fig.~\ref{fig:initial_configuration}(a)),
a dense-plate parallel to $yz$ plane (Fig.~\ref{fig:initial_configuration}(b)),
and a dense-cylinder parallel to $y$-axis (Fig.~\ref{fig:initial_configuration}(c)) 
under the identical set of parameters.
This initial velocity dependence appears in the region far from the coexistence phases,
where the system evolves from aggregates of many clusters (see Fig.~\ref{fig:dependence}).
%%%%%%%%%%%%%%%%%%%%%%%%%%%%%%
%%%%%%%%%%%%%%%%%%%%%%%%%%%%%%
%%%%%%%%%%%%%%%%%%%%%%%%%%%%%%
\subsection{Velocity distribution function}\label{sec:VDF}
\begin{figure}[htbp]
	\includegraphics[width=60mm]{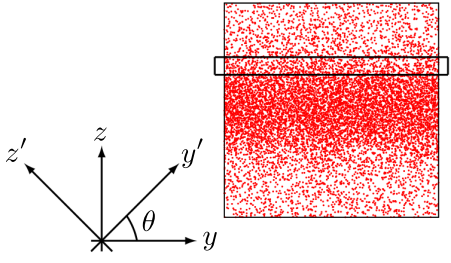}
	\caption{(Color online) A snapshot of our simulation for the plate-gases coexistence phase. 
Solid lines refer to the edges of a bin. The binwise velocity distribution function is calculated in each bin, whose width is $\Delta z=\sigma$. In addition, we introduce a new coordinate $(y^\prime, z^\prime)$, and $\theta$, which is the angle between $y^\prime$ and $y$-direction (in the counterclockwise direction) for later analysis.}
	\label{fig:binwise}
\end{figure}

\begin{figure}[htbp]
	\includegraphics[width=85mm]{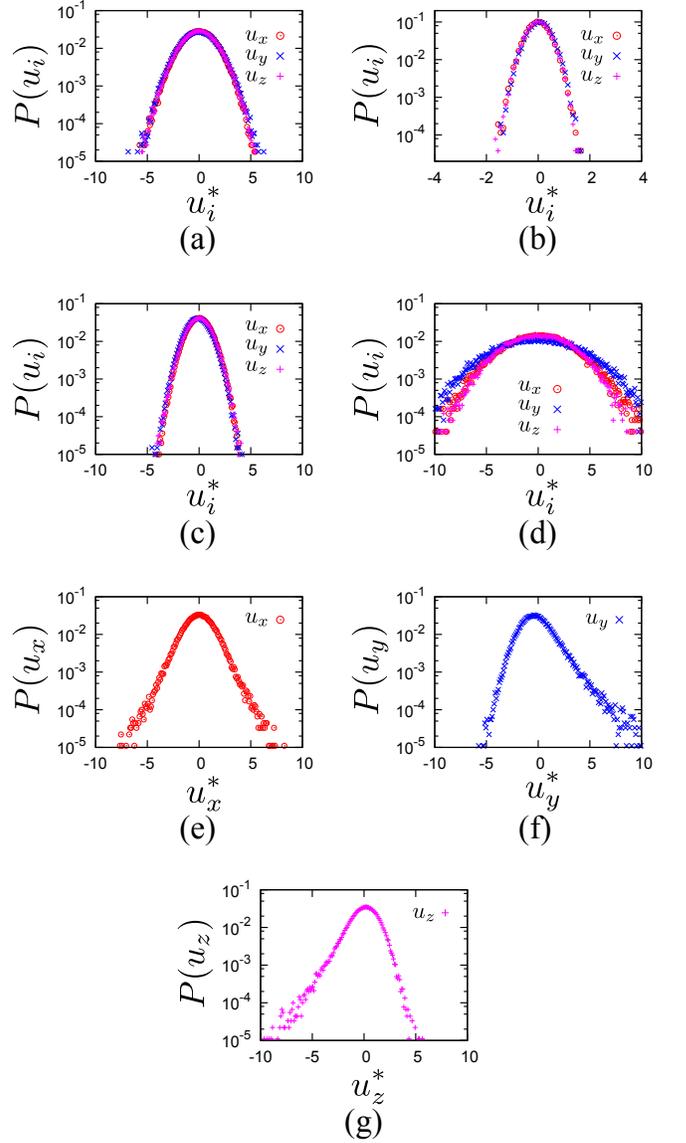}
	\caption{(Color online) Velocity distribution functions for various phases: 
			(a) VDFs in the phase Fig.~\ref{fig:steady}(a), 
			(b) VDFs in the phase Fig.~\ref{fig:steady}(g), 
			(c) VDFs in the dense region of the phase Fig.~\ref{fig:steady}(d),
			(d) VDFs in the dilute region of the phase Fig.~\ref{fig:steady}(d),
			(e) VDF of $x$-direction in the interface of the phase Fig.~\ref{fig:steady}(d),
			(f) VDF of $y$-direction in the interface of the phase Fig.~\ref{fig:steady}(d), and 
			(g) VDF of $z$-direction in the interface of the phase Fig.~\ref{fig:steady}(d).}
	\label{fig:VDF_interface}
\end{figure}

\begin{figure}[htbp]
	\includegraphics[width=86mm]{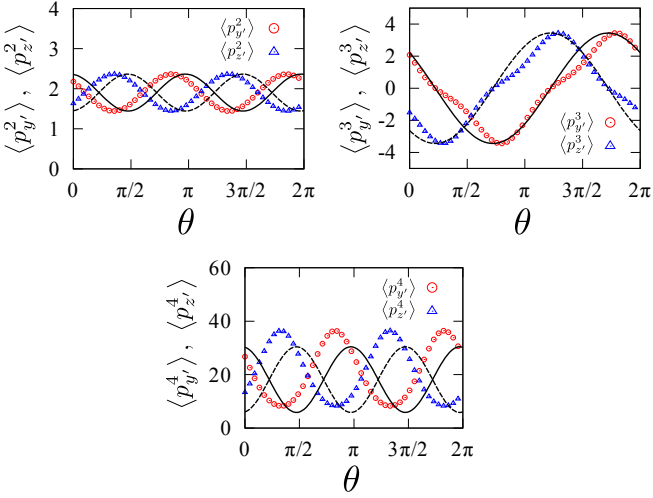}
	\caption{(Color online) The second, the third and the fourth moments obtained by MD 
	for $\rho^\ast=0.305$, $\dot\gamma^\ast=10^{-0.2}$, $\zeta^\ast=10^{0.2}$ 
	(circle: $y^\prime$-direction, upper triangle: $z^\prime$-direction) and those obtained by 
	Eqs.~(\ref{eq:2nd_y})--(\ref{eq:4th_z}) (solid line: $y^\prime$-direction, dashed line: $z^\prime$-direction).}
	\label{fig:moment}
\end{figure}

\begin{figure}[htbp]
	\includegraphics[width=85mm]{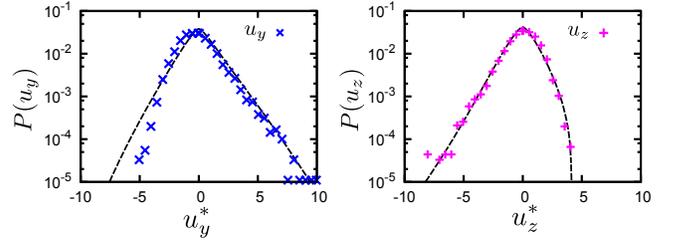}
	\caption{(Color online) VDFs in $y$-direction (left, cross) and $z$-direction (right, cross) obtained by our MD. The dashed lines in the left and right figures are the results of Eqs.~(\ref{eq:VDF_y}) and (\ref{eq:VDF_z}), respectively.}
	\label{fig:prob}
\end{figure}

%%%%%%%%%%%%%%%%%%%%%%%%%%%%%%
We also measure the velocity distribution function (VDF) $P(u_i)$ ($i=x,y,z$),
where $u_i$ is the velocity fluctuation around the mean velocity field, $\bar{v}_i$,
averaged over the time and different samples in the steady state.
For simplicity, we focus only on the following three phases;
the uniformly sheared phase (Fig.~\ref{fig:steady}(a)), 
the dense-plate coexistence phase (Fig.~\ref{fig:steady}(d)), 
and the dense-plate cluster phase (Fig.~\ref{fig:steady}(g)).
In this paper, we use the width $\Delta z=\sigma$ for bins in $z$-direction, while the bin sizes in both $x$ and $y$-directions are $L$ to evaluate VDF from our MD as in Fig.~\ref{fig:binwise}.
It is remarkable that the VDF is almost isotropic Gaussians for the phases corresponding to Figs.~\ref{fig:steady}(a) and (g)
as well as deep inside of both the dense and the gas regions in the coexistence phase 
in Fig.~\ref{fig:steady}(d) (see Figs.~\ref{fig:VDF_interface}(a)--(d)).
This is because we are interested in weak shear and weak dissipation cases without the influence of gravity.
On the other hand, VDF is nearly equal to an anisotropic exponential function \cite{Vijayakumar2007, Alam2010} 
in the vicinity of the interface between the dense and the gas regions in the coexistence phase corresponding to Fig.~\ref{fig:steady}(d)
as in Figs.~\ref{fig:VDF_interface}(e)--(g).
%%%%%%%%%%%%%%%%%%%%%%%%%%%%%%
%%%%%%%%%%%%%%%%%%%%%%%%%%%%%%
%%%%%%%%%%%%%%%%%%%%%%%%%%%%%%
%%%%%%%%%%%%%%%%%%%%%%%%%%%%%%
%%%%%%%%%%%%%%%%%%%%%%%%%%%%%%
\begin{figure}[htbp]
	\includegraphics[width=60mm]{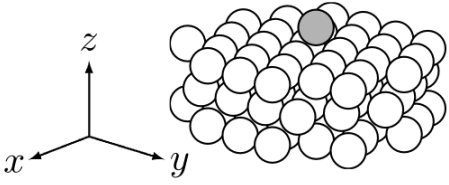}
	\caption{A schematic picture of the configuration of a gas particle (gray) and particles in the dense region (white).
			We assume that the wall particles are composed in a face-centered cubic lattice. We calculate the interaction energy between the gas particle and the wall particles whose distance is less than the cut-off length.}
	\label{fig:fcc}
\end{figure}
%%%%%%%%%%%%%%%%%%%%%%%%%%%%%%
%%%%%%%%%%%%%%%%%%%%%%%%%%%%%%
\begin{figure}[htbp]
	\includegraphics[width=70mm]{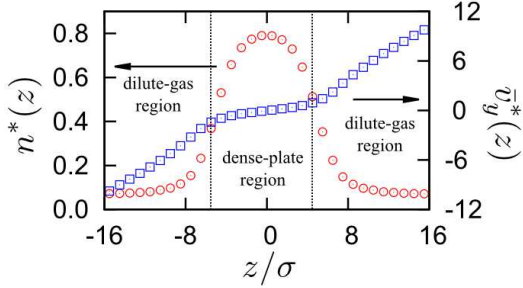}
	\caption{(Color online) The density and velocity profiles (in the $y$-direction) in the plate-gases coexistence phase ($n^*=0.305$, $\dot\gamma^*=10^{-0.2}$, $\zeta^*=10^{0.2}$),
	where $\bar{v}^*_y(z)=\bar{v}_y(z) \sqrt{m/\sigma\varepsilon}$.}
	\label{fig:profile}
\end{figure}
%%%%%%%%%%%%%%%%%%%%%%%%%%%%%%
\\
\quad
We now explain the non-Gaussian feature near the interface 
by a simple stochastic model of a tracer particle subjected to Coulombic friction
(the justification to use such a model is explained in Appendix \ref{sec:Coulombic}).
Let us consider a situation that a gas particle hits and slides on the wall formed by the particles in the dense region (see Fig.~\ref{fig:fcc}).
Because the velocity gradient in the gas region is almost constant as shown in Fig.~{\ref{fig:profile}},
we may assume that a tracer particle in the gas near the interface is affected by a plane shear.
Moreover, the tracer particle on a dense region may be influenced by Coulombic friction (see Appendix \ref{sec:Coulombic}).
When we assume that the collisional force among gas particles can be written as the Gaussian random noise $\xi$, 
the equations of motion of a tracer particle at the position $\bm{r}$ may be given by
\begin{align}
\frac{d\bm{r}}{dt} &=\frac{\bm{p}}{m}+\dot\gamma z\hat{\rm{e}}_y,\label{eq:EOM1}\\
\frac{d\bm{p}}{dt}&=-\mu F_0\frac{\bm{p}}{|\bm{p}|}-\dot\gamma p_z\hat{\bm{e}}_y+\bm{\xi},\label{eq:EOM2}
\end{align}
where $\bm{p}$ is a peculiar momentum, which is defined by Eq.~(\ref{eq:EOM1}).
Here we have introduced the friction constant $\mu_0$ and the effective force $F_0$ which is a function of the activation energy $\Delta E$ from the most stable trapped configuration of the solid crystal (see Fig.~\ref{fig:fcc}).
Here, $\xi$ is assumed to satisfy
\begin{align}
\left<\xi_\alpha(t)\right>=0,\quad \left<\xi_\alpha(t)\xi_\beta(t^\prime)\right>=2D\delta_{\alpha,\beta}\delta(t-t^\prime),
\end{align}
where $\left<\cdots\right>$ is the average over the distribution of the random variable $\bm{\xi}$. 
$D$ is the diffusion coefficient in the momentum space, which satisfies the fluctuation-dissipation relation $D=\mu F_0\sqrt{mT/(d+1)}$ in the $d$-dimensional system with a temperature $T$.
A set of Langevin equations (\ref{eq:EOM1}) and (\ref{eq:EOM2}) can be converted into the Kramers equation \cite{Kubo1991,Zwanzig2001,Kawarada2004, deGennes2005, Hayakawa2005}:
\begin{align}
\frac{\partial f}{\partial t}=&\left\{-\frac{\partial}{\partial \bm{r}}\cdot \left(\frac{\bm{p}}{m}+\dot\gamma z\hat{\bm{e}}_y\right) \right.\nonumber\\
&\left.+\frac{\partial}{\partial \bm{p}}\cdot \left(\dot\gamma p_z \hat{\bm{e}}_y +\mu F_0\frac{\bm{p}}{|\bm{p}|}+D\frac{\partial}{\partial \bm{p}}\right)\right\}f,\label{eq:FPeq}
\end{align}
where $f=f(\bm{r},\bm{p},t)$ is the probability distribution function of the tracer particle.
\\
\quad
If we multiply Eq.~(\ref{eq:FPeq}) by $p^2$ and integrate over $\bm{p}$, we immediately obtain
\begin{align}
\frac{\partial}{\partial t}\left<p^2\right>
=&-\frac{\partial}{\partial \bm{r}}\cdot \frac{\left<p^2\bm{p}\right>}{m}
-\dot\gamma z\frac{\partial}{\partial y}\left<p^2\right>\nonumber\\
&-2\dot\gamma\left<p_yp_z\right>-2\mu F_0\left<p\right>+2D,\label{eq:balance}
\end{align}
where $p=(p_y^2+p_z^2)^{1/2}$.
Because the third term on the right hand side (RHS) of Eq.~(\ref{eq:balance}) represents the viscous heating which is always positive as shown in Eq.~(\ref{eq:average_pypz}) and the fourth term is the loss of the energy due to friction,
the balance among the third, the fourth and the fifth terms on RHS of Eq.~(\ref{eq:balance}) produces a steady state.
It should be noted that the first and the second terms on RHS do not contribute to the energy balance equation for the whole system.
\\
\quad
Here, we only consider the steady distribution, i.e. $\partial f/\partial t=0$.
Thus, Eq.~(\ref{eq:FPeq}) is reduced to
\begin{align}
\frac{\bm{p}}{m}\cdot &\bm\nabla f+\dot\gamma z \frac{\partial}{\partial y}f-\dot\gamma p_z\frac{\partial}{\partial p_y}f\nonumber\\
&-\mu F_0 \frac{\partial}{\partial \bm{p}}\cdot \left(\frac{\bm{p}}{|\bm{p}|}f\right)
-D\Delta_{\bm{p}}f=0,\label{eq:exponential}
\end{align}
where $\Delta_{\bm{p}}=\partial^2/\partial p_y^2+\partial^2/\partial p_z^2$.
If there is neither a shear nor a density gradient, we find that Eq.~(\ref{eq:exponential}) has the steady solution obeying an exponential distribution, i.e. $f(\bm{p})= (\kappa^2/2\pi) \exp[-\kappa p]$,
where we have introduced $\kappa\equiv\mu F_0/D$.
We adopt the perturbative expression for $f$ in terms of $\epsilon\equiv \sigma/\lambda$, 
which is the ratio of the diameter $\sigma$ to the interface width $\lambda$, 
and the dimensionless shear rate $\dot\gamma^\ast$ as
(see the derivation in Appendix \ref{sec:perturbation})
\begin{align}
f(p,\theta)&= f^{(0,0)}(p,\theta) + \epsilon f^{(0,1)}(p,\theta)+\dot\gamma^\ast f^{(1,0)}(p,\theta).\label{eq:f_linear}
\end{align}
We also adopt the expansions 
\begin{align}
f^{(i,j)}(p,\theta)=\sum_{n=1}^\infty f_n^{(i,j)}\sin(n\theta),
\end{align}
with $(i,j)=(0,1)$ and $(1,0)$, where $\theta$ is the angle between $\bm{p}$ and $y$-axis (in the counterclockwise direction, see Fig.~\ref{fig:binwise}).
Then, we can solve Eq.~(\ref{eq:exponential}) perturbatively as
\begin{align}
f(p,\theta)&= f^{(0,0)}(p)+\epsilon f_1^{(0,1)}(p)\sin\theta + \dot\gamma^\ast f_2^{(1,0)}(p)\sin2\theta,\label{eq:f_per}
\end{align}
where $f^{(0,0)}$, $f_1^{(0,1)}$ and $f_2^{(1,0)}$ are, respectively, given by
\begin{align}
f^{(0,0)}(p)&=\frac{\kappa^2}{2\pi}\exp\left(-\kappa p\right),\\
f_1^{(0,1)}(p)&=-\frac{A}{6\pi\kappa}p\left(3+\kappa p+\kappa^2 p^2\right)\exp\left(-\kappa p\right),\\
f_2^{(1,0)}(p)&=-\frac{\kappa^2}{8\pi Dt_0}p^2\exp\left(-\kappa p\right).\label{eq:f_2nd}
\end{align}
Here, we have introduced $t_0=(m\sigma^2/\varepsilon)^{1/2}$ and $A$ given by Eq.~(\ref{eq:def_A}).
It should be noted that the other terms except for those in Eqs.~(\ref{eq:f_per})--(\ref{eq:f_2nd}) automatically disappear within the linear approximation as in Eq.~(\ref{eq:f_linear}).
\\
\quad
The second, the third and the fourth moments in $y^\prime$ and $z^\prime$-directions after the rotation by the angle of $\theta$ in the counterclockwise direction are, respectively, given by
\begin{align}
\left<p_{y^\prime,z^\prime}^2\right>&=\frac{3}{\kappa^2}\left(1 \mp \frac{5\dot\gamma}{2D\kappa^2}\sin2(\theta-\psi)\right),\label{eq:2nd_y}\\
\left<p_{y^\prime}^3\right>&=-\frac{765\epsilon A}{\kappa^6}\sin(\theta-\psi),\label{eq:3rd_y}\\
\left<p_{z^\prime}^3\right>&=-\frac{765\epsilon A}{\kappa^6}\cos(\theta-\psi),\label{eq:3rd_z}\\
\left<p_{y^\prime,z^\prime}^4\right>&=\frac{45}{\kappa^4}\left(1 \mp \frac{7\dot\gamma}{D\kappa^2}\sin2(\theta-\psi)\right),\label{eq:4th_z}
\end{align}
as shown in Appendix \ref{sec:moment}, where $\left<p_{y^\prime,z^\prime}^n\right>$ with $n=2$ or $4$ represents $\left<p_{y^\prime}^n\right>$ for a minus sign and $\left<p_{z^\prime}^n\right>$ for a plus sign, respectively.
To reproduce the node of the third moment in MD, we phenomenologically introduce the angle $\psi$ and replace $\theta$ by $\theta-\psi$ in Eqs.~(\ref{eq:2nd_y})--(\ref{eq:4th_z}).
Here, we choose $\psi=2\pi/9$ to fit the node position of the third moment.
We have not identified the reason why the direction of the node is deviated from the direction at which VDF becomes isotropic.
\\
\quad
Now, let us compare Eqs.~(\ref{eq:2nd_y})--(\ref{eq:4th_z}) with MD for a set of parameters $(n^\ast, \dot\gamma^\ast, \zeta^\ast)=(0.305,10^{-0.2},10^{0.2})$.
From the density profile (Fig.~\ref{fig:profile}) and the fitting to the second moment and the amplitude of the third moment,
we obtain $\epsilon\simeq 0.20$, $\mu\simeq1.3/\sqrt{m\varepsilon}$, $D=5.2\sqrt{m\varepsilon^3}/\sigma$, and $A\simeq0.088 /m^2\varepsilon^2$.
It is surprised that Eqs.~(\ref{eq:2nd_y})--(\ref{eq:4th_z}) can approximately reproduce the simulation results as in Fig.~\ref{fig:moment} except for the node positions of the second and the fourth moments.
%%%%%%%%%%%%%%%%%%%%%%%%%%%%%%
%%%%%%%%%%%%%%%%%%%%%%%%%%%%%%
\\
\quad
For the explicit form of VDF, at first, we convert $f(p,\theta)$ to $f(p_y, p_z)$ as in Appendix \ref{sec:rotation_f}:
\begin{align}
f(p_y,p_z)
=&\frac{\kappa^2}{2\pi}\exp\left(-\kappa p\right)\left[1+\frac{\epsilon A}{3\kappa^3}\left(3+\kappa p+\kappa^2 p^2\right)\right.\nonumber\\
&\left.\times(p_y\sin\psi-p_z\cos\psi)\right.\nonumber\\
&\left.+\frac{\dot\gamma}{4D}\left\{(p_y^2-p_z^2)\sin2\psi-2p_yp_z\cos2\psi\right\}\right].\label{eq:cartesian}
\end{align}
We obtain the peculiar velocity distribution function in each direction by integrating Eq.~(\ref{eq:cartesian}) with respect to $u_z$ or $u_y$ as
\begin{align}
P(u_y)=&\frac{m\kappa^2}{2\pi}\int_{-\infty}^\infty du_z 
\exp\left(-m\kappa u\right)\nonumber\\
&\times\left[1+\frac{m\epsilon A}{3\kappa^3}\left(3+m\kappa u+m^2\kappa^2u^2\right) u_y\sin\psi \right.\nonumber\\
&\left.+\frac{m^2\dot\gamma}{4D}(u_y^2-u_z^2)\sin2\psi\right]
,\label{eq:VDF_y}\\
P(u_z)=&\frac{m\kappa^2}{2\pi}\int_{-\infty}^\infty du_y 
\exp\left(-m\kappa u\right)\nonumber\\
&\times\left[1-\frac{m\epsilon A}{3\kappa^3}\left(3+m\kappa u+m^2\kappa^2u^2\right)u_z\cos\psi \right.\nonumber\\
&\left.+\frac{m^2\dot\gamma}{4D}(u_y^2-u_z^2)\sin2\psi\right]
,\label{eq:VDF_z}
\end{align}
where $u=(u_y^2+u_z^2)^{1/2}$.
These expressions semi-quantitatively reproduce VDF observed in our MD as in Fig.~\ref{fig:prob}.
%%%%%%%%%%%%%%%%%%%%%%%%%%%%%%
%%%%%%%%%%%%%%%%%%%%%%%%%%%%%%
%%%%%%%%%%%%%%%%%%%%%%%%%%%%%%
%%%%%%%%%%%%%%%%%%%%%%%%%%%%%%
%%%%%%%%%%%%%%%%%%%%%%%%%%%%%%
%%%%%%%%%%%%%%%%%%%%%%%%%%%%%%
%%%%%%%%%%%%%%%%%%%%%%%%%%%%%%
\section{discussion and conclusion}
\subsection{Discussion}
Let us discuss our results.
In Sec.~\ref{sec:Phase}, we do not discuss the time evolution of the granular temperature 
$T_{\rm g}=(m/3N)\sum_{i=1}^N |\bm{v}_i-\bm{V}|^2$,
where $\bm{V}=\bm{V}(\bm{r},t)$ is the ensemble average velocity field \cite{Goldhirsch2003, Goldhirsch2008}.
The granular temperature abruptly decreases to zero 
in the cluster phases Fig.~\ref{fig:steady}(e)--(i) when a big cluster which absorbs all gas particles appears \cite{Takada2013_2}.
To clarify the mechanism of abrupt change of the temperature during clusterings, we will need to study the more detailed dynamics.
\\
\quad
Moreover, to discuss the phase boundary between the uniformly sheared phase and the coexistence phases, 
we may use the stability analysis of a set of the hydrodynamic equations coupled with the phase transition dynamics \cite{Onuki2002}.
Once we establish the set of hydrodynamic equations, it is straightforward to perform weakly nonlinear analysis for this system \cite{Saitoh2011, Alam2013, Shukla2013}.
It should be noted that the set of equations may be only available
near the phase boundary between the uniformly sheared phase and the coexistence phases.
\\
\quad
In Fig.~\ref{fig:VDF_interface}, the VDF in a uniformly sheared phase is almost Gaussian.
This result seems to be inconsistent with the results for ordinary gases under a uniform shear flow \cite{Garzo2003}, which showed that the VDF differs from Gaussian even in a uniformly sheared phase.
In this study, however, we only restrict our interest to small inelastic and weakly sheared cases.
This situation validates small deviation from Gaussian.
%%%%%%%%%%%%%%%%%%%%%%%%%%%%%%
%%%%%%%%%%%%%%%%%%%%%%%%%%%%%%
%%%%%%%%%%%%%%%%%%%%%%%%%%%%%%
%%%%%%%%%%%%%%%%%%%%%%%%%%%%%%
\subsection{Conclusion}
We studied cohesive fine powders under a plane shear
by controlling the density, the dimensionless shear rate and the dissipation rate.
Depending on these parameters, 
we found the existence of various distinct steady phases as in Fig.~\ref{fig:steady}, 
and we have drawn the phase diagrams for several densities as in Fig.~\ref{fig:phasediagram}.
In addition, the shape of clusters depends on the initial condition of velocities of particles as in Fig.~\ref{fig:initial_configuration}, 
when the dissipation is strong.
We also found that there is a quasi particle-hole symmetry for the shape of clusters in steady states with respect to the density.
\\
\quad
We found that the velocity distribution functions near the interface between the dense region and the gas-like dilute region in the dense-plate coexistence phase deviate from the Gaussian as in Fig.~\ref{fig:VDF_interface}.
Introducing a stochastic model and its corresponding the Kramers equation (\ref{eq:FPeq}),
we obtain its perturbative VDFs as in Eqs.~(\ref{eq:VDF_y}) and (\ref{eq:VDF_z}),
which reproduce the semi-quantitative behavior of VDF observed in MD as in Fig.~\ref{fig:prob}.
This result suggests that the motion of a gas particle near the interface is subjected to Coulombic friction force whose origin is the activation energy in the dense region.
%%%%%%%%%%%%%%%%%%%%%%%%%%%%%%
%%%%%%%%%%%%%%%%%%%%%%%%%%%%%%
%%%%%%%%%%%%%%%%%%%%%%%%%%%%%%
\begin{acknowledgements}
The authors thank to Takahiro Hatano and Meheboob Alam for their fruitful advices.
KS wishes to express his sincere gratitude for the Yukawa Institute for Theoretical Physics (YITP) to support his stay and its warm hospitality.
Part of this work has been proceeded during YITP workshop YITP-WW-13-04 on ``Physics of Glassy and Granular Materials," YITP-T-13-03 on ``Physics of Granular Flow."
Numerical computation in this work was partially carried out at the Yukawa Institute Computer Facility.
This work is partially supported by Grant-in-Aid of MEXT, Japan (Grant No. 25287098).
\end{acknowledgements}
%%%%%%%%%%%%%%%%%%%%%%%%%%%%%%
%%%%%%%%%%%%%%%%%%%%%%%%%%%%%%
%%%%%%%%%%%%%%%%%%%%%%%%%%%%%%
\appendix
\section{Results of the physical boundary condition}\label{sec:LEbc}
\begin{figure}[htbp]
	\includegraphics[width=50mm]{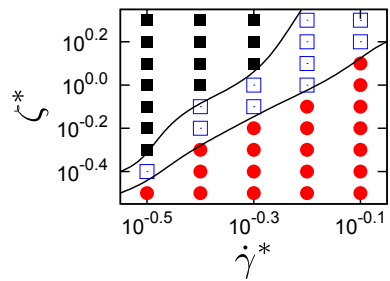}
	\caption{(Color online) Phase diagram under the flat boundary condition for $n^*=0.305$, 
	uniformly sheared state (red filled circle, Fig.~\ref{fig:steady}(a)), 
	coexistence of a dense-plate and gases (blue open square, Fig.~\ref{fig:steady}(d)), and 
	an isolated dense-plate (black filled square, Fig.~\ref{fig:steady}(g)).}
	\label{fig:physical}
\end{figure}

In this Appendix, we present the results of our simulations under the flat boundary condition which is one of the typical physical boundaries 
to clarify the influence of the boundary condition.
We prepare flat walls at $z=\pm L/2$, moving at velocities $\pm \dot\gamma L/2$ in $y$-direction, respectively.
When a particle with a velocity $(v_x, v_y, v_z)$ hits the walls at $z=\pm L/2$, 
the velocity is changed as $(v_x, \pm\dot\gamma L/2-v_y, -v_z)$ after the collision, respectively.
The phase diagram of the system for the physical boundary for $n^*=0.305$ is presented in Fig.~\ref{fig:physical}.
We have obtained three steady phases such as 
the uniformly sheared phase, 
the coexistence phase between dense-plate and gas regions, and 
the dense-plate cluster phase.
The phase diagram is almost same as the corresponding one
under the Lees-Edwards boundary condition (see Figs.~\ref{fig:phasediagram}(d)).
This can be understood as follows: 
if two particles at the symmetric positions with respect to the origin of the system 
simultaneously collide the walls at $z=L/2$ and $-L/2$, 
the pair of velocities after collisions is same 
as that after passing across the boundaries at $z=\pm L/2$ for the system under the Lees-Edwards boundary condition.
This is realized after the averaging over the collisions.
Thus, the flat boundary condition is essentially equivalent to the Lees-Edwards boundary condition.
%%%%%%%%%%%%%%%%%%%%%%%%%%%%%%
%%%%%%%%%%%%%%%%%%%%%%%%%%%%%%
%%%%%%%%%%%%%%%%%%%%%%%%%%%%%%
%%%%%%%%%%%%%%%%%%%%%%%%%%%%%%
\section{Calculation of Coulombic friction constant}\label{sec:Coulombic}
In this appendix, we try to illustrate the existence of Coulombic friction force for the motion of a tracer particle near the interface.
Let us consider a situation that a gas particle hits and slides on the wall formed by the particles in the dense region (see Fig.~\ref{fig:fcc}).
If the kinetic energy of the gas particle is less than the potential energy formed by the particles in the dense region,
it should be trapped in the potential well.
Therefore, the motion of the gas particle is restricted near the interface.
In this case, we can write the $N$-body distribution function near the interface $\rho(\bm{\Gamma},t)$ by using the distribution function in the equilibrium system as \cite{Evans2008, Chong2010_1, Chong2010_2, Hayakawa2013}
\begin{align}
\rho(\bm{\Gamma},t)=\rho_{\rm eq}(\bm{\Gamma})\exp\left[\int_0^t d\tau \Omega (-\tau, \bm{\Gamma}, \dot\gamma_l, \zeta)\right],\label{eq:Kawasaki2}
\end{align}
where $\bm{\Gamma}=\{\bm{r}_i, \bm{p}_i\}_{i=1}^N$, $\rho_{\rm eq}(\bm{\Gamma})$ is the equilibrium distribution function at time $t=0$, and
\begin{align}
%def of Omega
\Omega(t,\bm{\Gamma}, \dot\gamma, \zeta)
=&-\beta\dot\gamma V\sigma_{yz}(t,\bm{\Gamma},\dot\gamma,\zeta)\nonumber\\
&-2\beta {\cal R}(t,\bm{\Gamma},\dot\gamma,\zeta)-\Lambda(t,\bm{\Gamma},\dot\gamma,\zeta),
\end{align}
with
\begin{align}
%def of sigma
\sigma_{\alpha\beta}(t,\bm{\Gamma},\dot\gamma,\zeta)
=&\sum_{i}\left\{ \frac{p_{i,\alpha}p_{i,\beta}}{m} - \sum_{j\neq i} r_{i,\alpha}\frac{\partial U^{\rm LJ}(r_{ij})}{\partial r_{i,\beta}}\right.\nonumber\\
&\left.+ \sum_{j\neq i} r_{i,\alpha} F_{\beta}^{\rm vis}(\bm{r}_{ij},\bm{v}_{ij})\right\},\label{eq:def_sigma}\\
%def of R
{\cal R}(t,\bm{\Gamma},\dot\gamma,\zeta)
=&\frac{\zeta}{4}\sum_{i\neq j} \Theta(\sigma-r_{ij}) (\bm{v}_{ij}\cdot \hat{\bm{r}}_{ij})^2,\\
%def of Lambda
\Lambda(t,\bm{\Gamma},\dot\gamma,\zeta)
=&-\frac{\zeta}{m}\sum_{i\neq j}\Theta(\sigma-r_{ij}),\\
%def of F^vis
F_\beta^{\rm vis}(\bm{r}_{ij},\bm{v}_{ij})=&-\zeta\Theta(\sigma-r_{ij})(\bm{v}_{ij}\cdot \hat{\bm{r}}_{ij})\frac{r_{ij,\beta}}{r_{ij}}.
\end{align}
Here, we have introduced the inverse granular temperature $\beta=1/T$ and the local shear rate $\dot\gamma_l$ in the interface region.
If the dissipation is small and the shear rate is not large, we may assume that $\Omega(-t)\simeq -\beta \dot\gamma V\sigma_{yz}^{\rm mf}(-t)$, where $\sigma_{yz}^{\rm mf}$ is the mean field $yz$ component of the stress tensor.
We also assume that the stress tensor decays exponentially as $\sigma_{yz}^{\rm mf}(-t) \simeq \sigma_{yz}^{\rm mf} (0)\exp(-|t|/\tau_0)$ \cite{Evans2008}, where $\tau_0$ is the relaxation time of the stress tensor.
From these relationships, we may use the approximate expression
\begin{align}
\rho(\bm{\Gamma},t)\simeq& \prod_{i=1}^{N_l} \frac{1}{Z^{\rm mf}}\exp\left[-\beta \left(H^{\rm mf}-\Delta E_i\right)\right]\nonumber\\
&\times \exp\left(-\beta\tau_0\dot\gamma_l V_l\sigma_{yz}^{\rm mf}(0)\right),\label{eq:Kawasaki}
\end{align}
where $H^{\rm mf}$ and $\Delta E_i$, are respectively, the mean field Hamiltonian per particle in the interface and the energy fluctuation of the particle $i$ which may be the activation energy from the local trap.
Here $N_l$ and $V_l$ are, respectively, the number of particles and the volume in the interface region and $Z^{\rm mf}=\int d\bm{r}d\bm{p} \exp(-\beta H^{\rm mf})$.
There are two characteristic time scales $\dot\gamma^{-1}$ and $\dot\gamma_l^{-1}$ corresponding to the uniform region and the interface between dense and dilute regions.
Because the time scale is obtained from the average over the distribution function (\ref{eq:Kawasaki}) or the local mean field distribution, the relationship between $\dot\gamma^{-1}$ and $\dot\gamma_l^{-1}$ is expected to be
\begin{align}
\dot\gamma_l^{-1} = \dot\gamma^{-1} \exp\left[\beta(\Delta E- \tau_0\dot\gamma_l V_l\sigma_{yz}^{\rm mf}(0))\right],
\end{align}
where we have eliminated the suffix $i$ for the particle.
This equation can be rewritten as
\begin{align}
\sigma_{yz}^{\rm mf}(0)=\frac{1}{\tau_0\dot\gamma_l V_l}\left(\Delta E+T\ln \frac{\dot\gamma_l}{\dot\gamma}\right).
\end{align}
Therefore, we may estimate Coulombic friction constant as 
\begin{align}
\mu&=\frac{\sigma_{yz}^{\rm mf}(0)}{P}=\frac{1}{\tau_0\dot\gamma_l PV_l}\left(\Delta E+T\ln \frac{\dot\gamma_l}{\dot\gamma}\right),
\end{align}
where $P\simeq 0.90\varepsilon/\sigma^3$, $V_l\simeq4.3\sigma^3$, $\Delta E\simeq 3.5\varepsilon$ and $\dot\gamma_l\simeq 0.83(\varepsilon/m\sigma^2)^{1/2}$ at the interface for a set of parameters $(n^\ast,\dot\gamma^\ast, \zeta^\ast)=(0.305,10^{-0.2},10^{0.2})$.
In this expression, we cannot determine the relaxation time $\tau_0$ from the simulation,
which is estimated to reproduce the average value of the second moment with the aid of Eq.~(\ref{eq:2nd_y}).
%%%%%%%%%%%%%%%%%%%%%%%%%%%%%%
%%%%%%%%%%%%%%%%%%%%%%%%%%%%%%
%%%%%%%%%%%%%%%%%%%%%%%%%%%%%%
\section{Detailed calculation of the viscous heating term}\label{sec:av_pypz}
In this appendix, let us calculate the average of the viscous heating term by using the distribution function near the interface.
From Eq.~(\ref{eq:Kawasaki}), we can rewrite the distribution function with the aid of Eq.~(\ref{eq:def_sigma}) as
\begin{align}
\rho(\bm{\Gamma},t) \approx \frac{1}{Z} \prod_{i=1}^{N_l} \exp\left[-\beta\left(\frac{\bm{p}_i^2}{2m}+\tau_0 \dot\gamma_l V_l\frac{p_{i,y}p_{i,z}}{m}\right)\right],\label{eq:rho_eq}
\end{align}
where $Z=\int \prod_{i=1}^{N_l} d\bm{r}_i d\bm{p}_i \exp[-\beta (\bm{p}_i^2/2m+\tau_0\dot\gamma_l V_l p_{i,y}p_{i,z}/m)]$.
Then $\left<p_yp_z\right>$ is given by
\begin{align}
\left<p_yp_z\right>=& \int d\bm{\Gamma} p_{i,y}p_{i,z} \rho(\bm{\Gamma},t)\nonumber\\
\propto&\int_{-\infty}^\infty dp_{i,y}\int_{-\infty}^\infty dp_{i,z} p_{i,y}p_{i,z} \nonumber\\
&\times \exp\left[-\beta\left(\frac{\bm{p}_i^2}{2m}+\tau_0 \dot\gamma_l V_l\frac{p_{i,y}p_{i,z}}{m}\right)\right]\nonumber\\
=&\int_0^\infty dp \int_0^{2\pi}d\theta p^3 \sin\theta \cos\theta \nonumber\\
&\times \exp\left[-\beta \left(\frac{p^2}{2m}+\frac{\tau_0\dot\gamma_l V_l}{m}p^2\sin\theta \cos\theta\right)\right]\nonumber\\
=&-\frac{\pi}{2}\int_0^\infty dp p^3 \exp\left(-\frac{\beta p^2}{2m}\right) I_1\left(\frac{\beta \tau_0 \dot\gamma_l V_l}{2m}p^2\right),\label{eq:average_pypz}
\end{align}
where $I_1(x)$ is the modified Bessel function of the first kind \cite{Abramowitz1964}.
Because $I_1(x)$ is positive for $x>0$, Eq.~(\ref{eq:average_pypz}) ensures that the viscous heating term $-\dot\gamma\left<p_yp_z\right>$ is always positive near the interface.
%%%%%%%%%%%%%%%%%%%%%%%%%%%%%%
%%%%%%%%%%%%%%%%%%%%%%%%%%%%%%
%%%%%%%%%%%%%%%%%%%%%%%%%%%%%%
\section{A perturbative solution of the Kramers equation}\label{sec:perturbation}
In this appendix, let us solve the Kramers equation (\ref{eq:exponential}) perturbatively to obtain the steady VDF.
Later, we compare this solution with the result of MD.
\\
\quad
At first, we adopt the following three assumptions.
The first assumption is that the distribution function is independent of both $x$ and $y$, 
the coordinates horizontal to the interface.
We also assume that the distribution function $f$ depends on $z$, vertical to the interface, through the density and the granular temperature:
\begin{align}
\frac{\partial f}{\partial z}&=\frac{\partial f}{\partial n}\frac{dn}{dz}+\frac{\partial f}{\partial T}\frac{dT}{dz}.
\end{align}
Second, we assume that the changes of the density and the granular temperature near the interface can be characterized by the interface width $\lambda$ as
\begin{align}
\frac{dn}{dz}\simeq -\frac{n_0}{\lambda},\quad \frac{dT}{dz}\simeq \frac{T_0}{\lambda},
\end{align}
where $n_0=n(z_0)=(n_l+n_g)/2$, $T_0=T(z_0)=(T_l+T_g)/2$.
Here, $n_l$ and $T_l$ are the density and the granular temperature in the dense region, 
and $n_g$ and $T_g$ are those in the dilute region, respectively.
Third, we also assume that the interface width $\lambda$ is much longer than the diameter of the particles $\sigma$, i.e. $\epsilon\equiv \sigma / \lambda \ll 1$.
From these assumptions, $\partial f/\partial z$ may be rewritten as
\begin{align}
\frac{\partial f}{\partial z}&\simeq -\epsilon\left( \frac{n_0}{\sigma}\frac{\partial}{\partial n}
-\frac{T_0}{\sigma}\frac{\partial}{\partial T}\right)f.\label{eq:A0}
\end{align}
\\
\quad
To solve Eq.~(\ref{eq:exponential}), we adopt the perturbative expression Eq.~(\ref{eq:f_linear}).
Equation (\ref{eq:exponential}), thus, reduces to the following three equations:
for the zeroth order,
\begin{align}
-\kappa \frac{\partial}{\partial \bm{p}}\cdot \left(\frac{\bm{p}}{|\bm{p}|}f^{(0,0)}\right)-\Delta_{\bm{p}}f^{(0,0)}=0,\label{eq:0th}
\end{align}
for the first order of $\epsilon$,
\begin{align}
-\frac{p_z}{mD}&\left(\frac{n_0}{\sigma}\frac{\partial}{\partial n}
	-\frac{T_0}{\sigma}\frac{\partial}{\partial T}\right)f^{(0,0)}\nonumber\\
&-\kappa\frac{\partial}{\partial \bm{p}}\cdot \left(\frac{\bm{p}}{|\bm{p}|}f^{(0,1)}\right)
	-\Delta_{\bm{p}}f^{(0,1)}=0,\label{eq:1st01}
\end{align}
and for the first order of $\dot\gamma^\ast$,
\begin{align}
-\frac{p_z}{D}\frac{\partial f^{(0,0)}}{\partial p_y}
	-\kappa \frac{\partial}{\partial \bm{p}}\cdot \left(\frac{\bm{p}}{|\bm{p}|}f^{(1,0)}\right)
	-\Delta_{\bm{p}}f^{(1,0)}=0.\label{eq:1st10}
\end{align}
The solution of Eq.~(\ref{eq:0th}) is given by
\begin{align}
f^{(0,0)}&=C_1\exp(-\kappa p)+C_2\exp(-\kappa p){\rm Ei}\left(\kappa p\right),\label{eq:f_00}
\end{align}
where ${\rm Ei}(x)$ is the exponential integral ${\rm Ei}(x)\equiv -\int_{-x}^\infty (e^{-t}/t)dt$ \cite{Abramowitz1964},
and $C_1$ and $C_2$ are the normalization constants.
Here, we set $C_2=0$ because ${\rm Ei}(x)$ becomes infinite at $x=0$, 
and $C_1=\kappa^2/2\pi$ to satisfy the normalization condition without the shear and the density gradient.
Using Eq.~(\ref{eq:f_00}), Equations (\ref{eq:1st01}) and (\ref{eq:1st10}) can be represented in the polar coordinates as
\begin{align}
&A\left(p^2-\frac{2}{\lambda}p\right) f^{(0,0)}\sin\theta\nonumber\\
&=\kappa \left(\frac{1}{p}+\frac{\partial}{\partial p}\right)f^{(0,1)}+\left(\frac{\partial^2}{\partial p^2}
	+\frac{1}{p}\frac{\partial}{\partial p}
	+\frac{1}{p^2}\frac{\partial^2}{\partial \theta^2}\right)f^{(0,1)},\label{eq:f01}
\end{align}
and
\begin{align}
&\frac{\kappa}{2Dt_0}p f^{(0,0)}\sin2\theta\nonumber\\
&=\kappa\left(\frac{1}{p}+\frac{\partial}{\partial p}\right)f^{(1,0)}
	+\left(\frac{\partial^2}{\partial p^2}+\frac{1}{p}\frac{\partial}{\partial p}
	+\frac{1}{p^2}\frac{\partial^2}{\partial\theta^2}\right)f^{(1,0)},\label{eq:f10}
\end{align}
where we have introduced $A$ as
\begin{align}
A=\frac{n_0}{m\sigma D}\frac{\partial \kappa}{\partial n}
	-\frac{T_0}{m\sigma D}\frac{\partial \kappa}{\partial T}.\label{eq:def_A}
\end{align}
To solve Eqs.~(\ref{eq:f01}) and (\ref{eq:f10}), we adopt the expansions for $f^{(i,j)}(p,\theta)=\sum_{n=1}^\infty f_n^{(i,j)}(p)\sin(n\theta)$ with $(i,j)=(0,1)$ and $(1,0)$ \cite{Hayakawa2005}.
Equation (\ref{eq:f01}) for each $n$ reduces to the following equations:
for $n=1$,
\begin{align}
&\frac{A\kappa^2}{2\pi}\left(p^2-\frac{2}{\kappa}p\right)\exp(-\kappa p)\nonumber\\
&=\kappa\left(\frac{1}{p}+\frac{\partial}{\partial p}\right)f_1^{(0,1)}
	+\left(\frac{\partial^2}{\partial p^2}+\frac{1}{p}\frac{\partial}{\partial p}
	-\frac{1}{p^2}\right)f_1^{(0,1)},\label{eq:f01_1}
\end{align}
and for $n\neq 1$,
\begin{align}
0=\kappa\left(\frac{1}{p}+\frac{\partial}{\partial p}\right)f_n^{(0,1)}+\left(\frac{\partial^2}{\partial p^2}+\frac{1}{p}\frac{\partial}{\partial p}-\frac{n^2}{p^2}\right)f_n^{(0,1)}.\label{eq:f01_n}
\end{align}
The solutions of Eqs.~(\ref{eq:f01_1}) and (\ref{eq:f01_n}) are, respectively, given by
\begin{align}
f_1^{(0,1)}=&\frac{C_{11}}{p}+C_{12}\frac{1+\kappa p}{\kappa^2 p}\nonumber\\
	&-\frac{A}{6\pi}\frac{6+6\kappa p+3\kappa^2 p^2+\kappa^3p^3+\kappa^4p^4}{\kappa^3 p}\exp(-\kappa p),
\end{align}
and
\begin{align}
f_n^{(0,1)}=&C_{n1}(\kappa p)^n\exp(-\kappa p)U(n,2n+1,\kappa p)\nonumber\\
	&+C_{n2}(\kappa p)^n \exp(-\kappa p)L_{-n}^{2n}(\kappa p),
\end{align}
for $n\neq 1$, where $U(a,b,x)$ and $L_a^b(x)$ are, respectively, the confluent hypergeometric function and Laguerre's bi-polynomial \cite{Abramowitz1964}, and the normalization constants $C_{n1}$ and $C_{n2}$ ($n=1,2,\cdots$) will be determined later.
Similarly, Equation (\ref{eq:f10}) for each $n$ reduces to the following equations:
for $n=2$,
\begin{align}
&\frac{\kappa^3}{4\pi Dt_0}p\exp(-\kappa p)\nonumber\\
&=\kappa\left(\frac{1}{p}+\frac{\partial}{\partial p}\right)f_2^{(1,0)}
	+\left(\frac{\partial^2}{\partial p^2}+\frac{1}{p}\frac{\partial}{\partial p}
	-\frac{4}{p^2}\right)f_2^{(1,0)},\label{eq:f10_2}
\end{align}
and for $n\neq 2$,
\begin{align}
0=\kappa\left(\frac{1}{p}+\frac{\partial}{\partial p}\right)f_n^{(1,0)}
	+\left(\frac{\partial^2}{\partial p^2}+\frac{1}{p}\frac{\partial}{\partial p}
	-\frac{n^2}{p^2}\right)f_n^{(1,0)}.\label{eq:f10_n}
\end{align}
The solutions of Eqs.~(\ref{eq:f10_2}) and (\ref{eq:f10_n}) are, respectively, given by
\begin{align}
f_2^{(1,0)}=&C_{23}\frac{3-\kappa p}{p^2}
	+C_{24}\frac{6+4\kappa p+\kappa^2p^2}{\kappa^4 p^2}\exp(-\kappa p)\nonumber\\
	&+\frac{1}{8\pi Dt_0}\frac{72+48\kappa p+12\kappa^2p^2-\kappa^4p^4}{\kappa^2p^2}
	\exp(-\kappa p),
\end{align}
and
\begin{align}
f_n^{(1,0)}=&C_{n3}(\kappa p)^n\exp(-\kappa p)U(n,2n+1,\kappa p)\nonumber\\
	&+C_{n4}(\kappa p)^n \exp(-\kappa p)L_{-n}^{2n}(\kappa p),
\end{align}
for $n\neq2$, where the normalization constants $C_{n3}$ and $C_{n4}$ ($n=1,2,\cdots$) will be determined later.
\\
\quad
Here, let us determine the normalization constants $C_{n1}, \cdots, C_{n4}$ ($n=1,2,\cdots$).
The distributions $f_n^{(0,1)}$ and $f_n^{(1,0)}$ should be finite at $p=0$ and approach zero for large $p$.
Therefore, we obtain
\begin{align}
&C_{11}=0,\quad C_{12}=\frac{A}{\pi\kappa},\quad 
	C_{23}=0, \quad C_{24}=-\frac{3\kappa^2}{2\pi Dt_0},\nonumber\\
&C_{n1}=0,\quad C_{n2}=0 \quad (n\neq 1),\nonumber\\
&C_{n3}=0,\quad C_{n4}=0 \quad (n\neq 2).
\end{align}
From these results, we obtain
\begin{align}
f(p,\theta)&=f^{(0,0)}+\epsilon f_1^{(0,1)}\sin\theta + \dot\gamma^\ast f_2^{(1,0)}\sin2\theta,
\end{align}
where $f^{(0,0)}$, $f_1^{(0,1)}$ and $f_2^{(1,0)}$ are, respectively, given by
\begin{align}
f^{(0,0)}(p)&=\frac{\kappa^2}{2\pi}\exp(-\kappa p),\\
f_1^{(0,1)}(p)&=-\frac{A}{6\pi\kappa}p(3+\kappa p+\kappa^2 p^2)\exp(-\kappa p),\\
f_2^{(1,0)}(p)&=-\frac{\kappa^2}{8\pi Dt_0}p^2\exp(-\kappa p).
\end{align}
%%%%%%%%%%%%%%%%%%%%%%%%%%%%%%
%%%%%%%%%%%%%%%%%%%%%%%%%%%%%%
%%%%%%%%%%%%%%%%%%%%%%%%%%%%%%
\section{Detailed calculations of various moments}\label{sec:moment}
%%%%%%%%%%%%%%%%%%%%%%%%%%%%%%
In this appendix, we calculate the $n$-th moments of $p_{y^\prime}$ and $p_{z^\prime}$ using the distribution function obtained in Appendix \ref{sec:perturbation}.
From the definition of the moment, $n$-th moment of an arbitrary function $G(\bm{p})$ is given by
\begin{align}
\left<G^n\right>=\int d\bm{p} \hspace{0.2em} G^n(p,\varphi) f(p,\varphi).
\end{align}
We rotate the coordinate the coordinate $(y,z)$ by $\theta$ counterclockwise
and introduce the new Cartesian coordinate $(y^\prime, z^\prime)$ as in Fig.~\ref{fig:binwise}.
From this definition, we obtain the $n$-th moments of $p_{y^\prime}$,
for $n=2$,
\begin{align}
\left<p_{y^\prime}^2\right>
=&\int_0^\infty dp \int_0^{2\pi} d\varphi p^{3} \cos^2(\varphi-\theta) \nonumber\\
&\times\left[f^{(0,0)}(p)+\epsilon f_1^{(0,1)}(p)\sin\varphi + \dot\gamma^\ast f_2^{(1,0)}(p)\sin2\varphi\right]\nonumber\\
=&\frac{3}{\kappa^2}\left(1-\frac{5\dot\gamma}{2D\kappa^2}\sin2\theta\right),
\end{align}
for $n=3$,
\begin{align}
\left<p_{y^\prime}^3\right>
=&\int_0^\infty dp \int_0^{2\pi} d\varphi p^{4} \cos^3(\varphi-\theta) \nonumber\\
&\times\left[f^{(0,0)}(p)+\epsilon f_1^{(0,1)}(p)\sin\varphi + \dot\gamma^\ast f_2^{(1,0)}(p)\sin2\varphi\right]\nonumber\\
=&-\frac{765\epsilon A}{\kappa^7}\sin\theta,
\end{align}
and for $n=4$,
\begin{align}
\left<p_{y^\prime}^4\right>
=&\int_0^\infty dp \int_0^{2\pi} d\varphi p^{5} \cos^4(\varphi-\theta) \nonumber\\
&\times\left[f^{(0,0)}(p)+\epsilon f_1^{(0,1)}(p)\sin\varphi + \dot\gamma^\ast f_2^{(1,0)}(p)\sin2\varphi\right]\nonumber\\
=&\frac{45}{\kappa^4}\left(1-\frac{7\dot\gamma}{D\kappa^2}\sin2\theta\right).
\end{align}
Similarly, we can calculate the each moment of $p_{z^\prime}$ so that we obtain Eqs.~(\ref{eq:2nd_y})--(\ref{eq:4th_z}).
%%%%%%%%%%%%%%%%%%%%%%%%%%%%%%
%%%%%%%%%%%%%%%%%%%%%%%%%%%%%%
%%%%%%%%%%%%%%%%%%%%%%%%%%%%%%
\section{Velocity distribution function for each direction}\label{sec:rotation_f}
In this appendix, we derive the velocity distribution function in the Cartesian coordinate $(y,z)$ at first, and calculate the velocity distribution functions in $y$ and $z$-directions.
The velocity distribution function in the polar coordinates $(p,\theta)$ is given by Eq.~(\ref{eq:f_per}), 
where we replace $\theta$ by $\theta-\psi$ as in Eqs.~(\ref{eq:2nd_y})--(\ref{eq:4th_z}), 
which can be converted into the form in Cartesian coordinate as
\begin{widetext}
\begin{align}
f(p_y,p_z)
=&\frac{\kappa^2}{2\pi}\exp(-\kappa p)
	\left[1-\frac{\epsilon A}{3\kappa^3}p(3+\kappa p+\kappa^2p^2)\sin(\theta-\psi)
	-\frac{\dot\gamma}{4D}p^2\sin2(\theta-\psi)\right]\nonumber\\
=&\frac{\kappa^2}{2\pi}\exp(-\kappa p)
	\left[1+\frac{\epsilon A}{3\kappa^3}(3+\kappa p+\kappa^2p^2) (p_y\sin\psi-p_z\cos\psi)
	+\frac{\dot\gamma}{4D}\left\{(p_y^2-p_z^2)\sin2\psi-2p_yp_z\cos2\psi\right\}\right],\label{eq:f_pypz}
\end{align}
\end{widetext}
where $p=\sqrt{p_y^2+p_z^2}$.
Next, let us calculate the velocity distribution functions in $y$ and $z$ directions.
In this paper, we focus on the VDF for the fluctuation velocity, which is defined by the deviation from the average velocity.
Therefore, we can replace $p_y$ and $p_z$ by $mu_y$ and $mu_z$ in Eq.~(\ref{eq:f_pypz}).
The velocity distribution function in $y$-direction, $P(u_y)$, is given by integrating Eq.~(\ref{eq:f_pypz}) with respect to $u_z$ as
\begin{widetext}
\begin{align}
P(u_y)
=&\int_{-\infty}^\infty d(mu_z) f(mu_y,mu_z)\nonumber\\
=&\frac{m\kappa^2}{2\pi}\int_{-\infty}^\infty du_z \exp\left(-m\kappa u\right)
	\left[1+\frac{m\epsilon A}{3\kappa^3}\left(3+m\kappa u+m^2\kappa^2 u^2\right) u_y\sin\psi
	+\frac{m^2\dot\gamma}{4D}(u_y^2-u_z^2)\sin2\psi\right],
\end{align}
where $u=\sqrt{u_y^2+u_z^2}$.
Similarly, we can calculate the velocity distribution function in $z$-direction $P(p_z)$ as
\begin{align}
P(u_z)
=&\int_{-\infty}^\infty d(mu_y) f(mu_y,mu_z)\nonumber\\
=&\frac{m\kappa^2}{2\pi}\int_{-\infty}^\infty du_y \exp\left(-m\kappa u\right)
	\left[1-\frac{m\epsilon A}{3\kappa^3}\left(3+m\kappa u+m^2\kappa^2u^2\right)u_z\cos\psi
	+\frac{m^2\dot\gamma}{4D}(u_y^2-u_z^2)\sin2\psi\right].
\end{align}
\end{widetext}
%%%%%%%%%%%%%%%%%%%%%%%%%%%%%%
%%%%%%%%%%%%%%%%%%%%%%%%%%%%%%
%%%%%%%%%%%%%%%%%%%%%%%%%%%%%%
%%%%%%%%%%%%%%%%%%%%%%%%%%%%%%

\end{document}